\documentclass[9pt,twocolumn,twoside]{osajnl}

\journal{pr} 

\setboolean{shortarticle}{false}

\usepackage{lineno}
\usepackage{tabu, multirow}
\usepackage{bm,color}
\usepackage{xspace}
\usepackage[utf8]{inputenc}
\usepackage[T1]{fontenc}
\usepackage{mathptmx}
\usepackage{siunitx}
\usepackage{adjustbox}
\usepackage{diagbox}
\usepackage{booktabs}
\usepackage{xcolor,colortbl}
\definecolor{Gray}{gray}{0.85}
\usepackage{enumitem}
\usepackage{amsmath}
\usepackage{amssymb}

\makeatletter

\newcommand{\Rmnum}[1]{\expandafter\@slowromancap\romannumeral #1@}
\DeclareRobustCommand\onedot{\futurelet\@let@token\@onedot}

\def\@onedot{\ifx\@let@token.\else.\null\fi\xspace}
\def\ie{\emph{i.e}\onedot} 

\linenumbers

\title{10-mega pixel snapshot compressive imaging with a hybrid coded aperture}

\author[1,2,\dag]{Zhihong Zhang}
\author[1,2,\dag]{Chao Deng}
\author[3]{Yang Liu}
\author[4,*]{Xin Yuan}
\author[1,2,*]{Jinli Suo}
\author[1,2,5]{Qionghai Dai}

\affil[1]{Department of Automation, Tsinghua University, Beijing 100084, China}
\affil[2]{Institute for Brain and Cognitive Science, Tsinghua University, Beijing 100084, China}
\affil[3]{Computer Science and Artificial Intelligence Laboratory, Massachusetts Institute of Technology, Cambridge, Massachusetts 02139, USA}
\affil[4]{Bell Labs, 600 Mountain Avenue, Murray Hill, New Jersey 07974, USA}
\affil[5]{Beijing National Research Center for Information Science and Technology, Tsinghua University, Beijing 100084, China}
\affil[*]{Corresponding author: xyuan@bell-labs.com, jlsuo@tsinghua.edu.cn}

\begin{abstract}
High resolution images are widely used in our daily life, whereas high-speed video capture is challenging due to the low frame rate of cameras working at the high resolution mode.
Digging deeper, the main bottleneck lies in the {\bf \em{low throughput}} of existing imaging systems.  
Towards this end, snapshot compressive imaging (SCI) was proposed as a promising solution to improve the throughput of imaging systems by compressive sampling and computational reconstruction.
During acquisition, multiple high-speed images are encoded and collapsed to a single measurement. After this, algorithms are employed to retrieve the video frames from the coded snapshot. Recently developed Plug-and-Play (PnP) algorithms make it possible for SCI reconstruction in large-scale problems. However, the lack of high-resolution encoding systems still precludes SCI's wide application. In this paper, we build a novel hybrid coded aperture snapshot compressive imaging (HCA-SCI) system by incorporating a dynamic liquid crystal on silicon and a high-resolution lithography mask. We further implement a PnP reconstruction algorithm with cascaded denoisers for high quality reconstruction. Based on the proposed HCA-SCI system and algorithm, we achieve a 10-mega pixel SCI system to capture high-speed scenes, leading to a high throughput of 4.6G voxels per second. Both simulation and real data experiments verify the feasibility and performance of our proposed HCA-SCI scheme.
\end{abstract}

\setboolean{displaycopyright}{true}

\begin{document}

\maketitle

\section{Introduction}

\begin{figure}
\begin{center}
    \includegraphics[width=1.0\linewidth]{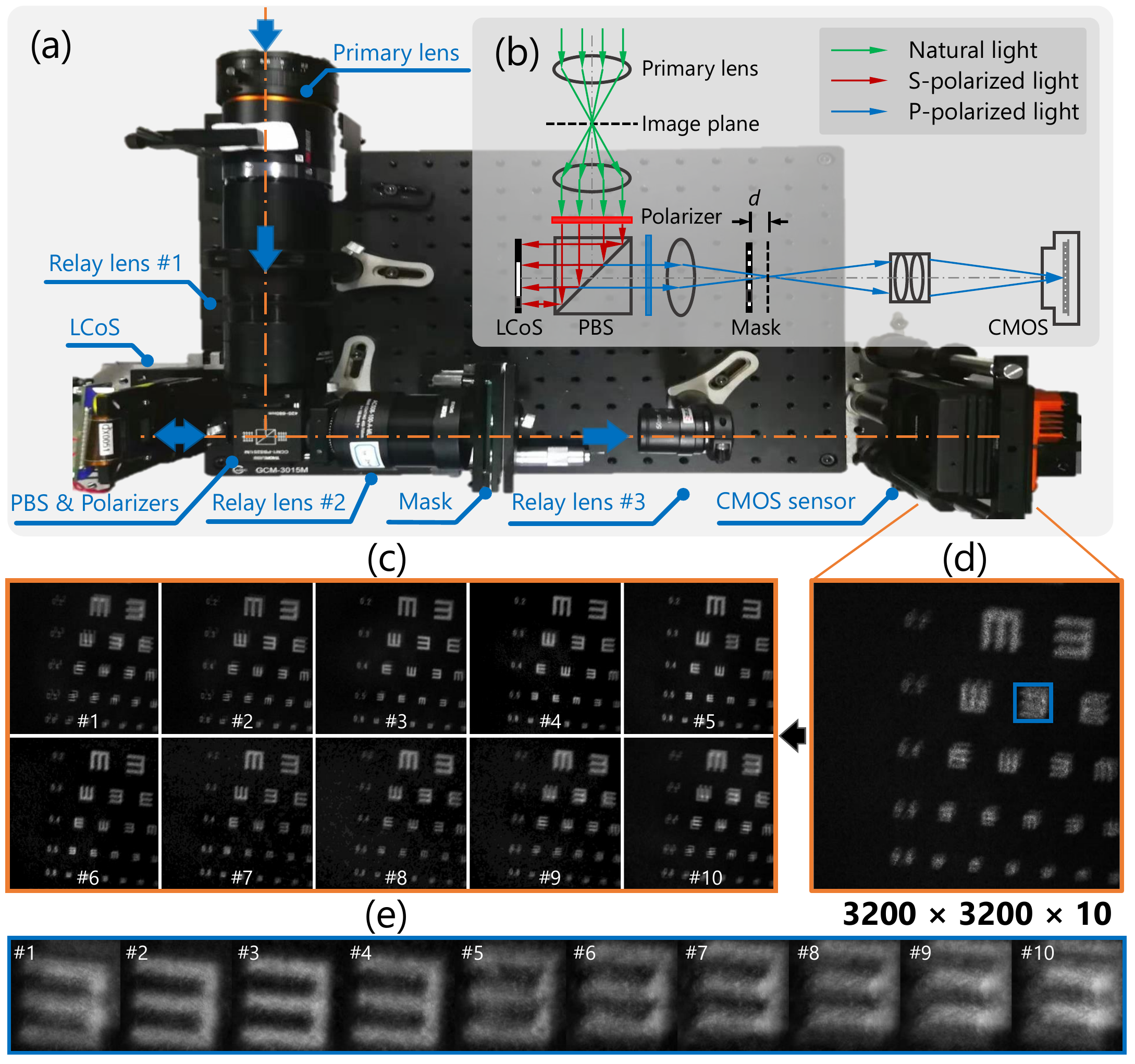}
\end{center}
\vspace{-3mm}
  \caption{Our 10-mega pixel video SCI system (a) and the schematic (b). 10 high-speed (200 fps) high-resolution (3200\texttimes 3200 pixels) video frames (c) reconstructed from a snapshot measurement (d), with motion detail in (e) for the small region in the blue box of (d). 
  Different from existing solutions that only uses an LCoS or a mask (thus with limited spatial resolution), our 10-mega spatio-temporal coding is jointly generated by an LCoS at the aperture plane and a static mask close to the image plane. }
\label{fig:real_cover}
\vspace{-3mm}
\end{figure}

Recent advances of machine vision with applications in robotics, drones, autonomous vehicles and cellphones have driven high resolution images into our daily life.
However, high-speed high-resolution videos though having wide applications in various fields such as physical phenomenon observation, biological fluorescence imaging, live broadcast of sports, on the other hand, are facing the challenge of {\em low throughput} due to the limited frame rate of existing cameras working at the high resolution mode. This is further limited by the memory, bandwidth  and power.
This paper aims to address this challenge by building a  high-speed high-resolution imaging system using compressive sensing.
Specifically, our system captures the high-speed scene in an encoded way and thus maintain the low bandwidth during capture. After this, reconstruction algorithms are employed to reconstruct the high speed high resolution scenes to achieve high throughput.
Note that though the idea of video compressive sensing has been proposed before, scaling it up to 10-mega pixel in spatial resolution presents the challenges of both hardware implementation and algorithm development. 
Please refer to {Fig.~\ref{fig:real_cover}} for a real high-speed scene captured by our newly built camera.

\begin{figure*}
\begin{center}
    \includegraphics[width=.96\linewidth]{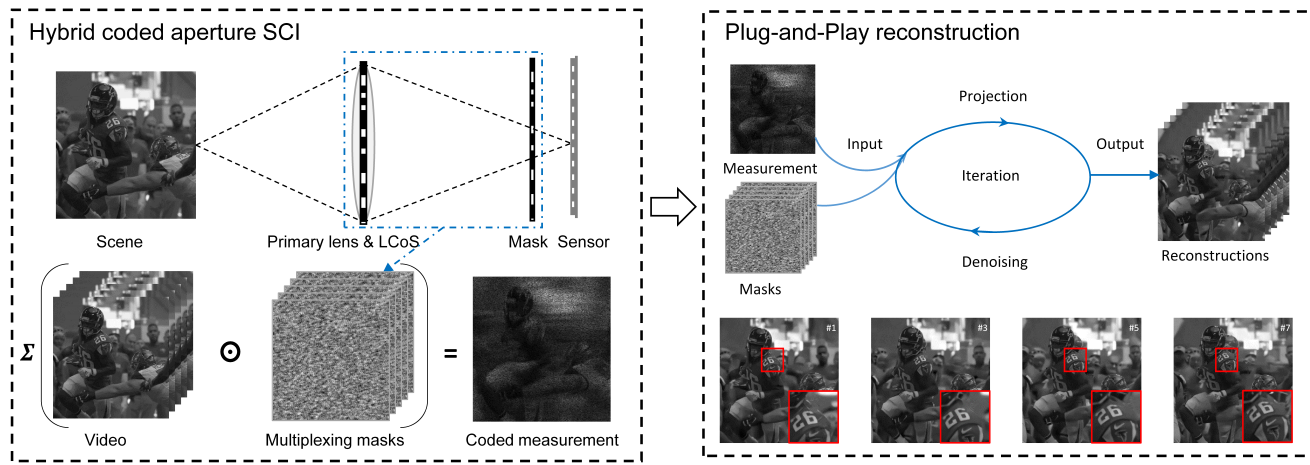}
\end{center}
\vspace{-3mm}
  \caption{The pipeline of the proposed large scale hybrid coded aperture snapshot compressive imaging (HCA-SCI) system (left) and the Plug-and-Play (PnP) reconstruction algorithms (right). {\bf Left}: During the encoded photography stage, a dynamic low resolution mask at the aperture plane and a static high resolution mask close to the sensor plane work together to generate a sequence of high resolution codes to encode the large scale video into a snapshot. {\bf Right}: In the decoding, the video is reconstructed under a PnP framework incorporating deep denoising prior into a  convex optimization (generalized alternating projection, GAP), which leverages the good convergence of GAP and the high efficiency of the deep network. }
\label{fig:overall}
\vspace{-3mm}
\end{figure*}

While 10-mega pixel lenses and sensors are both accessible, the main challenge for high-speed and high-resolution imaging lies in the deficient processing capability of current imaging systems. Massive data collected from high-speed high-resolution recording imposes dramatic pressure on the system’s storage and transmission modules, and thus makes it impossible for long-time capturing. In recent decades, the boosting of computational photography provides researchers with creative ideas and makes breakthroughs in many imaging-related fields like super-resolution \cite{carles2014superresolution, farsiu2004fast, marcia2008compressive}, deblurring \cite{cho2010motion,raskar2006coded,zhou2009what}, depth estimation \cite{zhou2011coded, sellent2014optimized, suwajanakorn2015depth} and so on. Regarding the {\em high throughput imaging}, snapshot compressive imaging (SCI) has been proposed and become a widely used framework \cite{llull2013coded,wagadarikar2008single,Yuan2021_SPM}. It aims to realize the reconstruction of high-dimensional data such as videos and hyper-spectral images from a single coded snapshot captured by a two-dimensional (2D) detector. A video SCI system is typically composed of an objective lens, a {\em temporally varying} mask, a monochrome or color sensor, and some extra relay lenses. During every single exposure, tens of temporal frames are modulated by corresponding temporal-variant masks and then integrated into a single snapshot. The high-dimensional data reconstruction in an SCI system can be formulated as an ill-posed linear model. Though different video SCI systems have been built~\cite{llull2013coded,reddy2011p2c2,hitomi2011video,qiao2020deep,yuan2014lowcost}, they are usually of low spatial resolution. By contrast, in this paper, we aim to build a {\em high resolution} video SCI system up to 10 mega pixels.

As mentioned above, the 10-mega pixel lenses (including imaging lens and relay lens) and sensors are both commercialized products. Off-the-shelf reconstruction algorithms such as the recently developed  plug-and-play  (PnP) framework~\cite{yuan2020plugandplay,Yuan2021_TPAMI} can also meet our demands in most real applications. However, the {\em 10-mega pixel  temporally varying mask} is still an open challenge. Classical SCI systems usually rely on shifting masks produced by the lithography technique or dynamic patterns projected by the spatial light modulator (SLM) (such as  the digital micromirror device (DMD) or liquid crystal on silicon (LCoS)) as temporally varying masks. The shifting mask scheme can provide high spatial resolution modulation, but it relies on the mechanical movement of the translation stage, which is inaccurate or unstable and can hardly be compact. For the masks generated by SLM or DMD, they can switch quickly with micro-mechanical controllers, but their resolution is generally limited to mega-pixel level, which is difficult to scale up. To the best of our knowledge, there are few SCI systems that can realize $1000 \times 1000$ pixel-resolution imaging in real-world scenes~\cite{sun2017compressive}. And typically, the resolution in prior works is mostly $256 \times 256$~\cite{llull2013coded} or $512 \times 512$~\cite{qiao2020deep}. 
Therefore, it is desirable to build a high-resolution video SCI system for real applications. 

To bridge this research gap, in this paper, we come up with a novel coded aperture imaging scheme, which leverages existing components to achieve the modulation up to 10-mega pixels. Our proposed modulation method is a hybrid approach using both a lithography mask and SLM. As depicted in Fig.~\ref{fig:overall}, during the encoding capture process, two modulation modules, an LCoS and a lithography mask are incorporated in different planes of the optical system. The LCoS with low spatial resolution is placed at the aperture plane of the imaging system, to dynamically encode the aperture and change the directions of incident lights. And the static lithography mask with high spatial resolution is placed in front of the image plane of the primary lens, which can project different high-resolution patterns on the image plane. When the LCoS changes its patterns, the lights propagating towards the lithography mask will change their directions accordingly, and thus leading to a different pattern. 
In this manner, we can implement dynamic modulation within one exposure time up to 10-mega pixels. Specifically, this paper makes the following contributions:
\begin{itemize}[leftmargin=*]
\setlength{\itemsep}{0pt}
\setlength{\parsep}{0pt}
\setlength{\parskip}{0pt}
    \item By jointly incorporating a dynamic LCoS and a high resolution lithography mask, we proposed a novel hybrid coded aperture snapshot compressive imaging (HCA-SCI) scheme, which can provide {\em multiplexed shifting patterns} to encode the image plane without physical movement of the mask.
    \item Inspired by the plug-and-play (PnP) algorithms for large-scale SCI in \cite{yuan2020plugandplay}, we implement a reconstruction algorithm which involves {\em cascading and series denoising} processes of total variation (TV) \cite{Rudin1992NTV} denoiser and learning-based FastDVDNet \cite{tassano2020fastdvdnet} denoiser. Simulation results show that the proposed algorithm can provide relative good reconstruction results in a reasonable time.
    \item Based on our proposed HCA-SCI scheme and the developed reconstruction algorithm, we build a 10-mega pixel large scale SCI system. Different compression rates of 6, 10, 20, 30 are implemented, providing reconstructed frame rate up to 450 frames per second (fps) for a conventional camera operating at 15 fps, verifying the effectiveness of proposed scheme in real scenarios.
\end{itemize}

\section{Related Work}
SCI has been proposed to capture high-dimensional data such as videos and hyper-spectral images from a single low-dimensional coded measurement. The underlying principle is to modulate the scene at a higher frequency than the frame rate of the camera and then the modulated frames are compressively sampled by a low-speed camera. Following this, inverse algorithms are employed to reconstruct the desired high dimensional data~\cite{Yuan2021_SPM}.  

Various video SCI systems have been developed in recent yeas \cite{deng2019sinusoidal, gao2014singleshot,hitomi2011video, llull2013coded, qiao2020deep, reddy2011p2c2,  waller2019video}, and the differences among these implementations mainly lie in the coding strategies. Typically, video SCI systems contain the same components as traditional imaging systems, except for several extra relay lenses and a modulation device that generates temporal-variant masks to encode the image plane. An intuitive approach is to directly use a DMD \cite{deng2019sinusoidal, qiao2020deep} or an LCoS \cite{reddy2011p2c2, hitomi2011video}, which can project given patterns with an assigned time sequence, on the image plane for image encoding. A substitute approach in early work is to simply replace the modulation device with a physically shifting lithography mask driven by a piezo \cite{llull2013coded}. There are also some indirect modulation methods proposed in recent work like \cite{waller2019video, gao2014singleshot}, which takes advantage of the temporal shifting feature of rolling shutter cameras or streak cameras for the temporal-variant mask generation. 

Paralleled to these systems, different algorithms are proposed to improve the SCI reconstruction performance as well. Since the inverse problem is ill-posed, different prior constrains such as TV \cite{yuan2016generalized}, sparsity~\cite{reddy2011p2c2, llull2013coded, yuan2014lowcost, hitomi2011video}, self-similarity\cite{yangliurank} and Gaussian mixture model (GMM) \cite{yang2014video, jianboyang2015compressive} are employed, 
forming widely used TwIST \cite{bioucas-dias2007new}, GAP-TV \cite{yuan2016generalized}, DeSCI \cite{yangliurank} and some other reconstruction algorithms. Generally, iterative optimization based algorithms have high computational complexity. Inspired by advances of deep learning, some learning-based reconstruction approaches are proposed and boost the reconstruction performance to a large extent \cite{cheng2020birnat,metasci,iliadis2018deep,kulkarni2016reconnet, madeep, qiao2020deep}. Recently, a sophisticated reconstruction algorithm BIRNAT \cite{cheng2020birnat} based on recurrent neural network (RNN) has led to state-of-the-art reconstruction performance with a significant reduction on the required time compared with DeSCI. However, despite the highest reconstruction quality achieved by learning-based methods, their main limitation lies in the inflexibility resulting from inevitable training process and requirement for large-scale training data when changing encoding masks or data capture environment. Other learning-based methods like MetaSCI~\cite{metasci} try to utilize meta-learning or transfer learning to realize fast mask adaption for SCI reconstruction with different masks, but the time cost is still unacceptable on the 10-mega pixel SCI data. To solve the trilemma of reconstruction quality, time consumption, and algorithm flexibility, a joint framework of iterative and learning-based methods called plug-and-play (PnP) \cite{yuan2020plugandplay} is proposed. By integrating pre-trained deep denoisers as the prior terms into certain iterative optimization process such as generalized alternating projection (GAP) \cite{liao2014generalized} and alternating direction method of multiplier (ADMM) \cite{boyd2011distributed}, PnP based approaches combine the advantages of both frameworks and realize the trade-off between speed, quality and flexibility.

In this paper, we build a novel video SCI system by using a hybrid coded aperture composed of an LCoS and a physical mask shown in Fig.~\ref{fig:real_cover}. Moreover, we modify the PnP algorithm to fit our system leading to better results than the method proposed in \cite{yuan2020plugandplay}. 

\section{System}

\subsection{Hardware implementation}
The hardware setup of our HCA-SCI system is depicted in Fig.~\ref{fig:real_cover}. It consists of a primary lens (HIKROBOT, MVL-LF5040M-F, $f$=50mm, $F$\#=4.0-32), an amplitude-modulated LCoS (ForthDD, QXGA-3DM, $2048\times1536$ pixels, 4.5k refresh rate), a lithography mask ($5120\times5120$ pixels, 4.5\textmu m\texttimes 4.5\textmu m pixel size), a CMOS camera (HIKROBOT, MV-CH250-20TC, $5120\times5120$ pixels, 4.5\textmu m pixel size), two achromatic doublets (Thorlabs, AC508-075-A-ML, $f$=75mm), a relay lens (ZLKC, HM5018MP3, $f$=50mm), a polarizing beamsplitter (Thorlabs, CCM1-PBS251/M), and two film polarizers (Thorlabs, LPVISE100-A). The incident light from a scene is first collected by the primary lens and focused at the first virtual image plane. Then a 4f system consisting of two achromatic doublets transfers the image through the aperture coding module and the lithography mask, and subsequently onto the second virtual image plane. The aperture coding module positioned at the middle of the 4f system is composed of a polarizing beamsplitter, two film polarizers and an amplitude-modulated LCoS, which are used to change the open-close states (`open' means letting the light go through while `close' means blocking the light) of the sub-apertures and thus modulate the light's propagation direction. Finally, the image delivered by the 4f system is relayed to the camera sensor being captured. Note that the 4f system used in our system has a magnification of 1, and the relay lens has a magnification of 2, which on the whole provides a 1:2 mapping between pixels of the lithography mask and the sensor. Even though the imaging model involves pixel-wise encoding, there is no need for a precise alignment for the lithography mask and the camera sensor, since the actual encoding mask will be calibrated before data acquisition. During the acquisition process, the camera shutter is synchronized with the LCoS by using an output trigger signal from the LCoS driver board.

It is worth nothing that, the active area of the LCoS and the position of the lithography mask should be carefully adjusted. The active area of the LCoS should be as large as possible meanwhile to ensure it to serve as the aperture stop of the whole system, so that it can provide a higher light efficiency. As for the lithography mask, fine-tune is needed to ensure that the mask's projection on the image plane can generate a shifting when different parts of the aperture is open, and meanwhile, the shifting masks can still keep sharp. In our implementation, after extensive experiments, the mask is placed in front of the second image plane with a distance of 80$\mu$m, which is a good trade-off between the sharpness and the shift.

\subsection{Encoding mask generation}
\label{sec: encoding_mask_gen}
\begin{figure}[t]
\begin{center}
  \includegraphics[width=\linewidth]{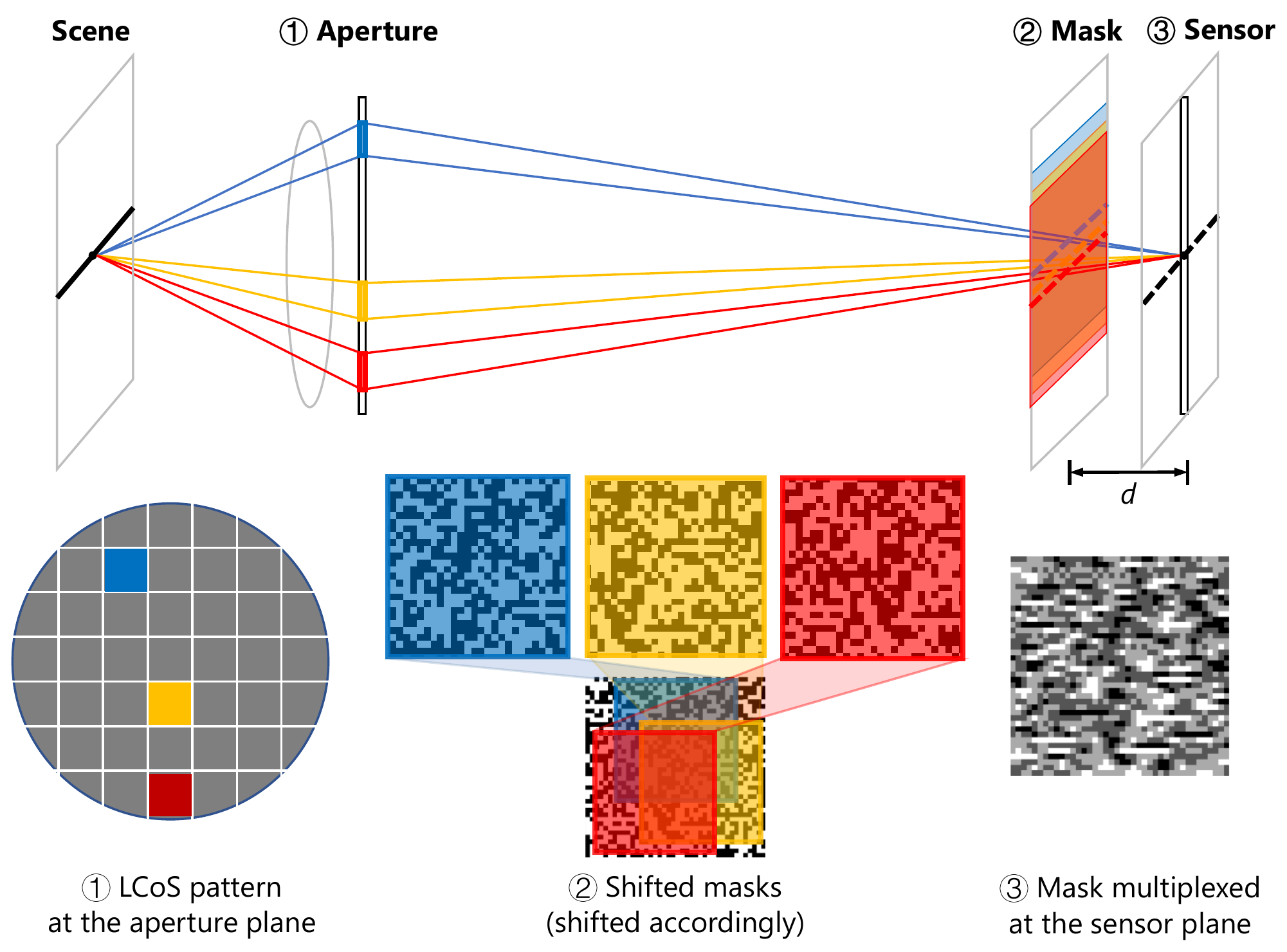}
\end{center}
\vspace{-2mm}
  \caption{The illustration of the multiplexed mask generation. For the same scene point, its images generated by different sub-apertures (marked as blue, yellow and red) intersect the mask plane with different regions and thus encoded with corresponding (shifted) random masks before summation at the sensor. The multiplexing would raise the light flux for high signal-to-noise-ratio recording, while with only slight performance degeneration.} 
\label{fig:mask_gen}
\vspace{-2mm}
\end{figure}

The aperture of the system (\ie the activate area of the LCoS) can be divided into several sub-apertures according to the resolution of the LCoS after pixel binning, and each sub-aperture corresponds to a light beam propagating towards certain directions. As shown in Fig.~\ref{fig:mask_gen}, because the lithography mask is placed in front of the image plane, when different sub-apertures are turned on, the light beams from the corresponding sub-apertures will project the mask onto different parts of the image plane, which can thus generate corresponding shifting encoding masks. In practice, to enhance the light throughput, multiple sub-apertures will be turned on simultaneously in one frame by assigning the LCoS with a specific multiplexing pattern to obtain a single multiplexing encoding mask. And in different frames, different combinations of the sub-apertures are applied to generate different multiplexing encoding masks. Generally, we turn on 50\% of the sub-apertures in one multiplexing pattern. In this multiplexing case, the final encoding mask on the image plane will be the summation of those shifting masks provided by the corresponding sub-apertures.

\subsection{Mathematical model}
\label{sec: math_model}

Mathematically, the encoding mask generation process can be modeled as a multiplexing of shifting masks. Let $\bm{O}$ denote the center-view mask generated by opening the central sub-aperture of the system, then each mask $\bm{C}$ generated by sub-aperture multiplexing can be formulated as
\begin{equation}
  \textstyle  \bm{C} =  \sum_{i=1}^{N}m_{i} {\rm S}_i(\bm{O}),
    \label{eq:multiplex}
\end{equation}
where ${\rm S}_i$ donates the mask shifting operator corresponding to the \textit{i}-th sub-aperture; $m_i$ is the multiplexing indicator (a scalar) for the \textit{i}-th sub-aperture, with 0 and 1 for blocking or transmitting the light respectively; and $N$ is the amount of sub-apertures (i.e. the number of macro-pixels in the active area of the LCoS after binning).
Consider that a video $\bm{X} \in \mathbb{R}^{n_x \times n_y \times B}$, containing $B$ consecutive high speed frames is modulated by $B$ encoding masks $\bm{C} \in \mathbb{R}^{n_x \times n_y \times B}$ and integrated by the sensor to generate a snapshot coded measurement $\bm{Y}$. Then $\bm{Y}$ can be expressed as 
\begin{equation}
    \textstyle \bm{Y} = \sum_{k=1}^{B} \bm{C}_k \odot \bm{X}_k+\bm{G},
    \label{eq:sci}
\end{equation}
where $\odot$ denotes the Hadamard (element-wise) product; $\bm{G} \in \mathbb{R}^{n_x \times n_y}$ is the measurement noise; $\bm{C}_k = \bm{C}(:,:,k)$ and $\bm{X}_k = \bm{X}(:,:,k)$ represent the \textit{k}-th multiplexed mask and corresponding frame respectively. Through a simple derivation, the coded measurement in \eqref{eq:sci} can be further expressed by
\begin{equation}
    \bm{y} = \bm{H}\bm{x} + \bm{g},
    \label{eq:sci_mat}
\end{equation}
where $\bm{y} = {\rm Vec}(\bm{Y}) \in \mathbb{R}^n$ and $\bm{g} = {\rm Vec}(\bm{G}) \in \mathbb{R}^n$ with $n = n_x n_y$. Corresponding video signal $\bm{x} \in \mathbb{R}^{nB}$ is given by
    $\bm{x} = [\bm{x}_1^\top, \ldots, \bm{x}_B^\top]^\top$,
where $\bm{x}_k = {\rm Vec}(\bm{X}_k) \in \mathbb{R}^n$.
Different with traditional CS, the coding matrix $\bm{H} \in \mathbb{R}^{n \times nB}$ in video SCI is sparse and has a special structure which can be written as
\begin{equation}
    \bm{H} = [\bm{D}_1, \ldots, \bm{D}_B],
    \label{eq:sci_mat_H}
\end{equation}
where $\bm{D}_k = {\rm diag}( {\rm Vec} (\bm{C}_b)) \in  \mathbb{R}^{n \times n}$. Therefore, the compressive sampling rate in video SCI is equal to $1/B$. Theoretical analysis in \cite{jalali2018compressivea, jalali2019snapshota} has proved that the reconstruction error of SCI is bounded even when $B>1$.

\subsection{system calibration}
\label{sec:sys_calib}
Though pixel-wise modulation is involved in SCI systems, there is no need for pixel to pixel alignment during system building, as we can get the precise encoding patterns through an "end-to-end" calibration process before data capture. To be specific, we place a Lambertian white board at the objective plane, and provisionally take away the lithography mask. Then an illumination pattern $\bm{I}$ and a background pattern $\bm{B}$ are captured with LCoS projecting white and black patterns, respectively. After that, we put on the lithography mask and capture each dynamic encoding mask $\bm{C}$ with LCoS projecting corresponding multiplexing patterns directly.To eliminate the influence of background light and nonuniform illumination caused by light source or system vignetting on the actual encoding masks, we conduct calibration to get the accurate encoding mask $\bm{C'}$ following \eqref{eq:calib}. Besides, after encoding acquisition, the encoded measurements will also subtract the background pattern to accord with the providing mathematical model, and illumination will be regarded as a part of the scene itself.

\begin{equation}
    \bm{C'} =\frac{\bm{C}-\bm{B}}{\bm{I}-\bm{B}},
    \label{eq:calib}
\end{equation}

\section{Reconstruction Algorithm}
The reconstruction of high speed videos from the snapshot coded measurement is an ill-posed problem. As mentioned before, to solve this problem, different priors and frameworks have been employed. 
Roughly, the algorithms can be categorized into the following 3 classes~\cite{Yuan2021_SPM}: $i$) Regularization (or priors) based optimization algorithms with well known methods such as TwIST~\cite{bioucas-dias2007new}, GAP-TV~\cite{yuan2016generalized} and DeSCI~\cite{yangliurank}. $ii$) End-to-end deep learning based algorithms~\cite{qiao2020deep,iliadis2018deep,madeep}, like BIRNAT~\cite{cheng2020birnat} which reaches state-of-the-art performance and recent developed MetaSCI~\cite{metasci} that using meta-learning to improve adaption capability for different masks in SCI reconstruction.
$iii$) Plug-and-Play (PnP) algorithms that using deep denoising networks into the optimization framework such as ADMM and GAP. 

Among these algorithms, regularization based algorithms are usually too slow and end-to-end deep learning networks need a large amount of data and also a long time to train the network, in addition to the inflexibility,  \ie, re-training being required for a new system. Though recent works like MetaSCI~\cite{metasci} try to mitigate this problem with meta-learning or transfer learning and thus march forward to large-scale SCI problem with patch-wise reconstruction strategy, it still takes a long time for the training and adaption of 10-meta pixel scale SCI reconstruction. For example, MetaSCI takes about two weeks for the $256\times256\times10$ base model training on a single NVIDIA 2080Ti GPU, and further adaption performed on more than 570 $256\times256\times10$ sub-tasks (overlapped patches extracted from a 10-mega pixel image) costs about two months, which is impractical in real applications (more GPUs can be used to mitigate this challenge). By contrast, PnP has achieved a well balance of speed, flexibility and accuracy. 
Therefore, we employ the PnP framework in our work and further develop the PnP-TV-FastDVDNet to achieve promising results for our high-resolution HCA-SCI scheme.
Meanwhile, we use GAP-TV and BIRNAT as baselines for comparison purpose. 

In the following, we review the main steps of PnP-GAP~\cite{yuan2020plugandplay} and then present our PnP-TV-FastDVDNet algorithm for HCA-SCI in Algorithm \ref{algo:PnP_GAP}.

\subsection{PnP-GAP} \label{sec:PnP-GAP}
In GAP, the SCI inversion problem is modeled as
\begin{equation}
  \textstyle   \mathop{\arg\min}_{\bm{x}, \bm{v}} \ \frac{1}{2}\| \bm{x}-\bm{v}\|_2^2 + \lambda R(\bm{x}),\ \mathrm{s.t.} \ \bm{y} = \bm{H}\bm{x},
    \label{eq:gap}
\end{equation}
where $R(\bm{x})$ is a regulizer or prior being imposed on $\bm{x}$, which can be a TV, sparse prior or a deep prior~\cite{zhang2018ffdnet} and $\bm{v}$ is an auxiliary parameter.
Let $k$ index the iteration number, through a two-step iteration, the minimization in \eqref{eq:gap} can be solved as follows:
\begin{itemize}[leftmargin=*]
\setlength{\itemsep}{0pt}
\setlength{\parsep}{0pt}
\setlength{\parskip}{0pt}
    \item Solving $\bm{x}$:
    \begin{equation}
        \bm{x}^{(k)} = \bm{v}^{(k-1)}+\bm{H}^\top(\bm{H}\bm{H}^\top)^{-1}(\bm{y}-\bm{H}\bm{v}^{(k-1)}),
        \label{eq:gap-x}
    \end{equation} 
    By utilizing the special structure of $\bm{H}$ shown in \eqref{eq:sci_mat_H}, this subproblem can be solved efficiently via element-wise operation rather than calculating the inversion of a huge matrix.
    \item Solving $\bm{v}$:
    \begin{equation}
        \bm{v}^{(k)} = \mathcal{D}_{\sigma}(\bm{x}^{(k)}).
        \label{eq:gap-v}
    \end{equation}  
    where $\mathcal{D}_{\sigma}(\cdot)$ represents a denoising process with $\sigma = \sqrt{\lambda}$. Here, different denoising algorithms can be used such as TV (thus GAP-TV), WNNM~\cite{gu2014weighted} (thus DeSCI~\cite{yangliurank}) and FFDnet~\cite{zhang2018ffdnet} (thus PnP-FFDnet~\cite{yuan2020plugandplay}).
\end{itemize}

\subsection{PnP-TV-FastDVDNet}
Recall that solving $\bm{v}$ following \eqref{eq:gap-v} is equivalent to performing a denoising process on $\bm{x}$. By plugging various denoisers into the GAP iteration steps, we can make a trade-off between different aspects of reconstruction performance. In fact, more than one denoiser can be employed in a series manner, i.e. one after another in each iteration; or in a cascading manner, i.e. the first several iterations using one denoiser, while the next several iterations using another. In this way, we can further balance the strengths and drawbacks of different denoisers.

\begin{algorithm}[!htbp]
	\caption{PnP-TV-FastDVDNet for HCA-SCI}
	\begin{algorithmic}[1]
		\Require$\bm{H},\bm{y}$.
		\State {Initialize:} $\bm{v}^{(0)}, \lambda_0, \xi<1, k=1, K_1, K_{Max}$.
		\While{Not Converge \textbf{and} $k\le K_{Max}$ }
		\State Update $\bm{x}$ by \eqref{eq:gap-x} .
		\State Update $\bm{v}$:
		\If {$k \le K_1$}
		\State {$\bm{v}^{(k)} = \mathcal{D}_{TV}(\bm{x}^{(k)})$}
		\Else 
		\State {$\bm{v}' = \mathcal{D}_{TV}(\bm{x}^{(k)})$}
		\State {$\bm{v}^{(k)} = \mathcal{D}_{FastDVDNet}(\bm{v}')$}
		\EndIf
		\EndWhile
	\end{algorithmic}
	\label{algo:PnP_GAP}
\end{algorithm}

\begin{figure}[hbp]
\begin{center}
  \includegraphics[width=\linewidth]{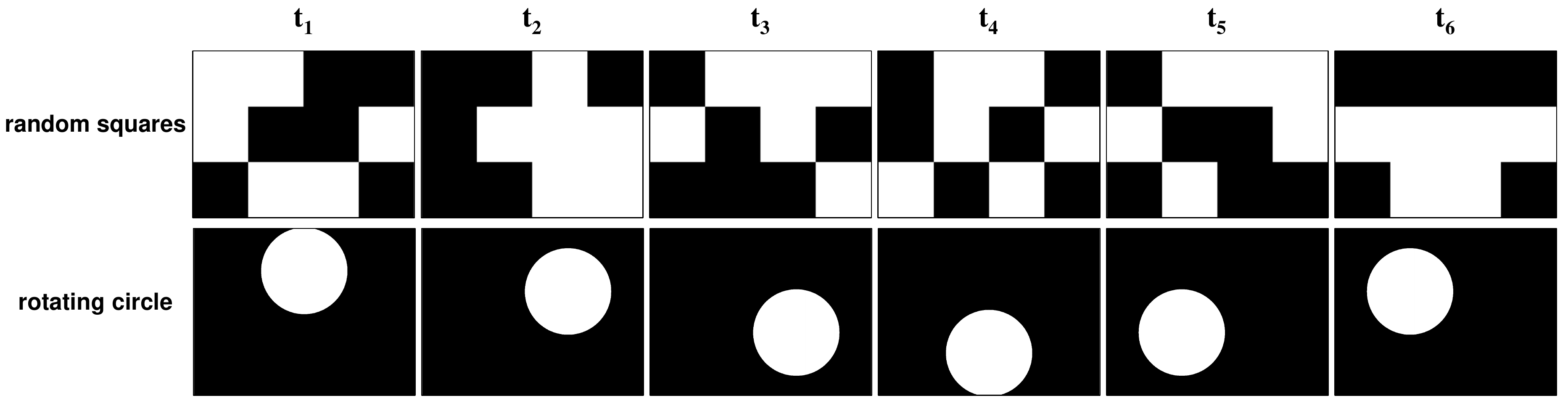}
\end{center}
\vspace{-2mm}
  \caption{Multiplexing pattern schemes used in our experiments (taking Cr = 6 for an example). {\bf Top row}: multiplexing patterns for simulation experiments. Each pattern contains 50\% "open" sub-apertures, and each sub-apertures is a 512$\times$512 binning macro pixel on the LCoS. {\bf Bottom row}: multiplexing patterns for real experiments. Each pattern contains an "open" circle with a radius of about 400 pixels, and the circles in adjacent patterns have a rotation of 360/Cr degrees.}
\label{fig:multiplex_pattern}
\vspace{-2mm}
\end{figure}

\begin{table*}[htbp!]
\caption{The average results of PSNR in dB (left entry in each cell) and SSIM (right entry in each cell) by different algorithms (Cr = 10). BIRNAT fails at large-scale due to limited GPU memory.}
\vspace{-3mm}
\begin{center}
\resizebox{1.0\textwidth}{!}{
    \begin{tabular}{c|c|ccccc|c}
        \hline
        Scales & Algorithms & \texttt{Football} & \texttt{Hummingbird} & \texttt{ReadySteadyGo} & \texttt{Jockey} & \texttt{YachtRide} & Average \\
        \hline \hline
        \multirow{4}{*}{$256 \times 256$} & GAP-TV & 27.82, 0.8280 & 29.24, 0.7918 & 23.73, 0.7499 & 31.63, 0.8712 & 26.65, 0.8056 & 27.81, 0.8093  \\
        & PnP-FFDNet & 27.06, 0.8264 & 25.52, 0.6912 & 21.68, 0.6859 & 31.14, 0.8493 & 23.69, 0.7035 & 25.82, 0.7513 \\ 
        &  \textbf{PnP-TV-FastDVDNet} & \textbf{31.31, 0.9123} & \textbf{31.19, 0.8264} & \textbf{26.18, 0.8276} & \textbf{31.36, 0.8817} & \textbf{28.90, 0.8841} & \textbf{29.79, 0.8664}   \\
        & BIRNAT & 34.67, 0.9719 & 34.33, 0.9546 & 29.50, 0.9389 & 36.24, 0.9711 & 31.02, 0.9431 & 33.15, 0.9559  \\
        \hline
        \multirow{4}{*}{$512 \times 512$} & GAP-TV & 29.19, 0.8854 & 28.32, 0.7887 & 25.94, 0.7918 & 31.30, 0.8718 & 26.59, 0.7939 & 28.27, 0.8263  \\
        & PnP-FFDNet & 28.57, 0.8952 & 28.02, 0.8363 & 24.32, 0.7457 & 29.81, 0.8248 & 23.45, 0.6793 & 26.83, 0.7963   \\   
        & \textbf{PnP-TV-FastDVDNet} & \textbf{30.92, 0.9333} & \textbf{32.24, 0.8834} & \textbf{27.04, 0.8246} & \textbf{32.11, 0.8839} & \textbf{27.87, 0.8487} & \textbf{30.04, 0.8748} \\
        \hline
        \multirow{4}{*}{$1024 \times 1024$} & GAP-TV & 30.63, 0.9022 & 29.16, 0.8459 & 28.92, 0.8698 & 31.59, 0.8953 & 29.03, 0.8470 & 29.87, 0.8720  \\
        & PnP-FFDNet & 29.87, 0.9023 & 27.70, 0.7869 & 27.70, 0.8483 & 29.88, 0.8412 & 25.55, 0.7211 & 28.14, 0.8200 \\
        &  \textbf{PnP-TV-FastDVDNet} & \textbf{30.35, 0.9265} & \textbf{31.71, 0.8909} & \textbf{29.42, 0.8913} & \textbf{31.59, 0.9014} & \textbf{30.44, 0.8713} & \textbf{30.70, 0.8963} \\ 
        \hline
        \hline
    \end{tabular}
}
\end{center}
\label{tb:exp1_Cr10}
\vspace{-5mm}
\end{table*}

\begin{table*}[htbp!]
\caption{The average results of PSNR in dB (left entry in each cell) and SSIM (right entry in each cell) by different algorithms (Cr = 20). BIRNAT fails at large-scale due to limited GPU memory.}
\vspace{-3mm}
\begin{center}
\resizebox{1.0\textwidth}{!}{
\begin{tabular}{c|c|ccccc|c}
    \hline
    Scales & Algorithms & \texttt{Football} & \texttt{Hummingbird} & \texttt{ReadySteadyGo} & \texttt{Jockey} & \texttt{YachtRide} & Average \\
    \hline \hline
    \multirow{4}{*}{$256 \times 256$} & GAP-TV & 25.01, 0.7544 & 26.33, 0.6893 & 20.48, 0.6326 & 28.13, 0.8318 & 23.56, 0.7129 & 24.70, 0.7242  \\
    & PnP-FFDNet & 21.67, 0.6657 & 22.13, 0.5835 & 17.27, 0.5340 & 27.78, 0.7994 & 20.39, 0.6024 & 21.85, 0.6370   \\   
    &  \textbf{PnP-TV-FastDVDNet} & \textbf{27.83, 0.8459} & \textbf{28.65, 0.7520} & \textbf{23.28, 0.7381} & \textbf{29.51, 0.8597} & \textbf{26.34, 0.8235} & \textbf{27.12, 0.8038} \\
    & BIRNAT & 27.91, 0.9021 & 28.58, 0.8800 & 23.79, 0.8279 & 31.35, 0.9467 & 26.14, 0.8585 & 27.55, 0.8830\\
    \hline 
    \multirow{4}{*}{$512 \times 512$} & GAP-TV & 23.97, 0.8179 & 24.50, 0.6719 & 22.12, 0.6975 & 26.99, 0.8297 & 23.13, 0.6930 & 24.14, 0.7420  \\
    & PnP-FFDNet & 22.00, 0.7661 & 23.62, 0.7245 & 19.35, 0.6133 & 25.32, 0.7924 & 19.48, 0.5418 & 21.95, 0.6876   \\ 
    &  \textbf{PnP-TV-FastDVDNet} & \textbf{25.63, 0.8852} & \textbf{28.36, 0.7778} & \textbf{23.80, 0.7499} & \textbf{28.79, 0.8553} & \textbf{25.36, 0.7784} & \textbf{26.39, 0.8093}  \\
    \hline
    \multirow{4}{*}{$1024 \times 1024$} & GAP-TV & 24.82, 0.8353 & 25.53, 0.7296 & 24.98, 0.8128 & 26.63, 0.8388 & 25.80, 0.7759 & 25.55, 0.7985 \\
    & PnP-FFDNet & 23.55, 0.8098 & 23.02 0.6039 & 22.48, 0.7702 & 24.48, 0.7968 & 21.67, 0.6414 & 23.04, 0.7244   \\ 
    & \textbf{PnP-TV-FastDVDNet} & \textbf{26.26, 0.8729} & \textbf{28.68, 0.8076} & \textbf{26.31, 0.8399} & \textbf{29.18, 0.8773} & \textbf{28.07, 0.8194} & \textbf{27.70, 0.8434} \\
    \hline
    \hline
\end{tabular}
}
\end{center}
\label{tb:exp1_Cr20}
\vspace{-3mm}
\end{table*}

\begin{figure*}[htbp!]
\begin{center}
    \includegraphics[width=1\linewidth]{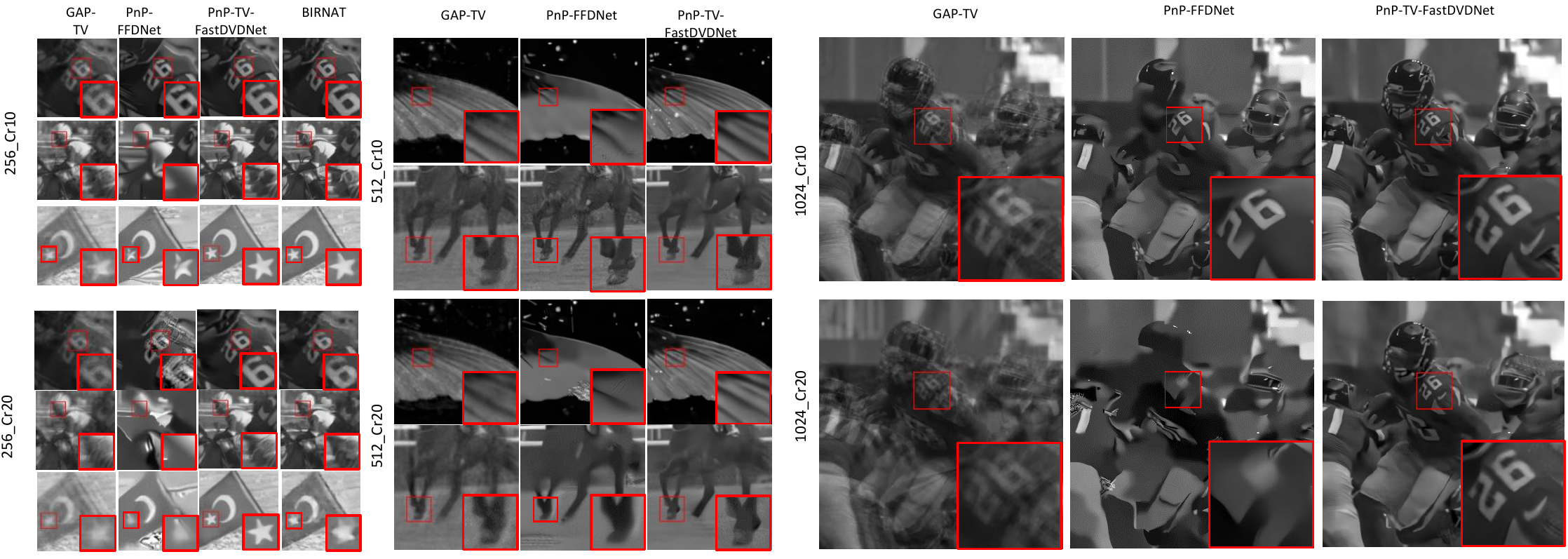}
\end{center}
\vspace{-4mm}
  \caption{Reconstruction results and comparison with state-of-the-art algorithms on simulated data at different resolutions (left: 256\texttimes 256, middle: 512\texttimes 512, right: 1024\texttimes 1024) and with different compression ratios (top: Cr = 10, bottom: Cr = 20). The BIRNAT results are not available for 512\texttimes 512 and 1024\texttimes 1024 since the model training will be our of memory. See Visualization 1-6 for the reconstructed videos.}
\label{fig:exp1}
\vspace{-4mm}
\end{figure*}
In this paper, considering the high video denoising performance of recently proposed FastDVDnet~\cite{tassano2020fastdvdnet}, we jointly employ the TV denoiser and the FastDVDNet denoiser in a PnP-GAP framework, implementing a reconstruction algorithm of PnP-GAP-TV+FastDVDNet, which involves cascading and series denoising processes. The algorithm pipeline is shown in Algorithm \ref{algo:PnP_GAP}. In each iteration, the updating of $\bm{x}$ follows \eqref{eq:gap-x}, and the updating of $\bm{v}$, i.e., the denoising process, differs in different periods. To be specific, in the first period, a single TV denoiser is employed, while in the second period, a joint TV and FastDVDNet denoiser is involved one after another.

\section{Results}
In this section, we conduct a series of experiments on both simulation data and real data to validate the feasibility and performance of our proposed HCA-SCI system. Four reconstruction algorithms including iterative optimization based algorithm GAP-TV~\cite{yuan2016generalized}, plug-and-play based algorithm PnP-FFDNet~\cite{yuan2020plugandplay}, our proposed PnP-TV-FastDVDNet, and the state-of-the-art learning-based algorithm BIRNAT~\cite{cheng2020birnat} are employed.

\subsection{Multiplexing pattern design}
\label{multiplex_patt}
Multiplexing pattern design plays an important role in our HCA-SCI system. In simulation, we used the “random squares” multiplexing scheme (shown in the first row of Fig.~\ref{fig:multiplex_pattern}), which contains 12 sub-apertures with each sub-aperture containing 512\texttimes512 binning pixels. In each multiplexing pattern, there are 50\% randomly selected sub-apertures being "open". In real experiments, the optical aberration and diffraction that are not considered in simulation experiments will cause blur to the encoding mask, and thus introduce more correlation among the encoding masks which degrades the encoding effect. Considering the fact that the calibration process is conducted in an "end-to-end" manner in real experiments, we thus have enough degree of freedom to design any possible multiplexing patterns without being limited by the binning mode or the sub-aperture amount. We empirically tried many different schemes, and finally found that the “rotating circle” scheme (shown in the second row of Fig.~\ref{fig:multiplex_pattern}) can realized a good balance between the system's light throughput and the encoding masks’ quality when taking the physical factors into account.

Currently, the design of the  multiplexing patterns is heuristic, and the challenge of the algorithm-based multiplexing pattern design mainly lies in the large size of the multiplexing pattern (1536$\times$2048) and the video data (10-mega) in our system. Both traditional optimization-based methods and learning based methods have difficulty in dealing with data of this scale. So, the design of the optimal multiplexing pattern is still an open challenge which is worthy for future investigation.

\subsection{Reconstruction comparison between different algorithms on simulation datasets}

In order to investigate the reconstruction performance of different algorithms on the proposed HCA-SCI system, we first perform experiments on simulated datasets, which involve three different scales of $256 \times 256$, $512 \times 512$, $1024 \times 1024$ and two compressive rates (Cr) of 10 and 20. Two datasets named \texttt{Football} and \texttt{Hummingbird} used in \cite{yuan2020plugandplay} and three datasets named \texttt{ReadySteadyGo}, \texttt{Jockey} and \texttt{YachtRide} provided in \cite{mercat2020uvg} are employed in our simulation. According to the forward model depicted in Section \ref{sec: math_model}, to simulate the encoding masks, we first calculate the mask shifting distance with respect to the center-view mask for each sub-aperture in the current multiplexing pattern (shown in the first row of Fig.~\ref{fig:multiplex_pattern}) based on geometry optics. Then we shift the center-view mask accordingly to get each sub-aperture’s shifting mask. And finally the shifting masks are added together to obtain the final multiplexing encoding mask on the image plane. After that, we can generate six groups of simulation datasets by modulating different scales of videos (containing 10 or 20 frames) with multiplexed shifting masks generated from the HCA-SCI system and then collapsing the coded frames to a single (coded) measurement. 

The reconstruction PSNR and SSIM for each algorithm are summarized in Table \ref{tb:exp1_Cr10} and Table  \ref{tb:exp1_Cr20} for Cr = 10 and 20, respectively. It is worth nothing that due to the limited memory of our GPU (GeForce RTX 3090 with 24GB memory), we only test BIRNAT on $256 \times 256$ scale datasets with Cr equaling to 10 and 20, and the adversarial training is not involved. For GAP-TV and PnP-TV-FastDVDNet, the reconstruction is conducted on a platform equipped with a Intel(R) Core(TM) i7-9700K CPU (3.60GHz, 32G memory) and a GeForce RTX 2080 GPU with 8G memory; 160 and 250 iterations are taken for Cr = 10 and 20, respectively. It can be observed that: $i$) On all six groups of simulated datasets, our proposed PnP-TV-FastDVDNet outperforms GAP-TV and PnP-FFDNet for about 1.5dB and 4dB on average, respectively. 
$ii$) On the \texttt{256\_Cr10} and \texttt{256\_Cr20} datasets, BIRNAT leads the best performance  (with sufficient training) exceeding PnP-TV-FastDVDNet for about 3.4dB and 0.4dB, respectively. 
$iii$) BIRNAT is the fastest algorithm during inference period with hundreds times shorter time than GAP-TV and PnP-TV-FastDVDNet. However, the training process of BIRNAT is quite time-consuming, and it takes about one week to train the network without adversarial training for 25 epochs. 
$iv$) From the selected reconstruction video frames in Fig.~\ref{fig:exp1}, we can see that our proposed PnP-TV-FastDVDnet provides higher visualization quality with sharp edges and less artifacts. By contrast, GAP-TV produces noisy results while PnP-FFDnet leads to some unpleasant artifacts. 

\begin{figure}[htbp!]
\begin{center}
    \includegraphics[width=.9\linewidth]{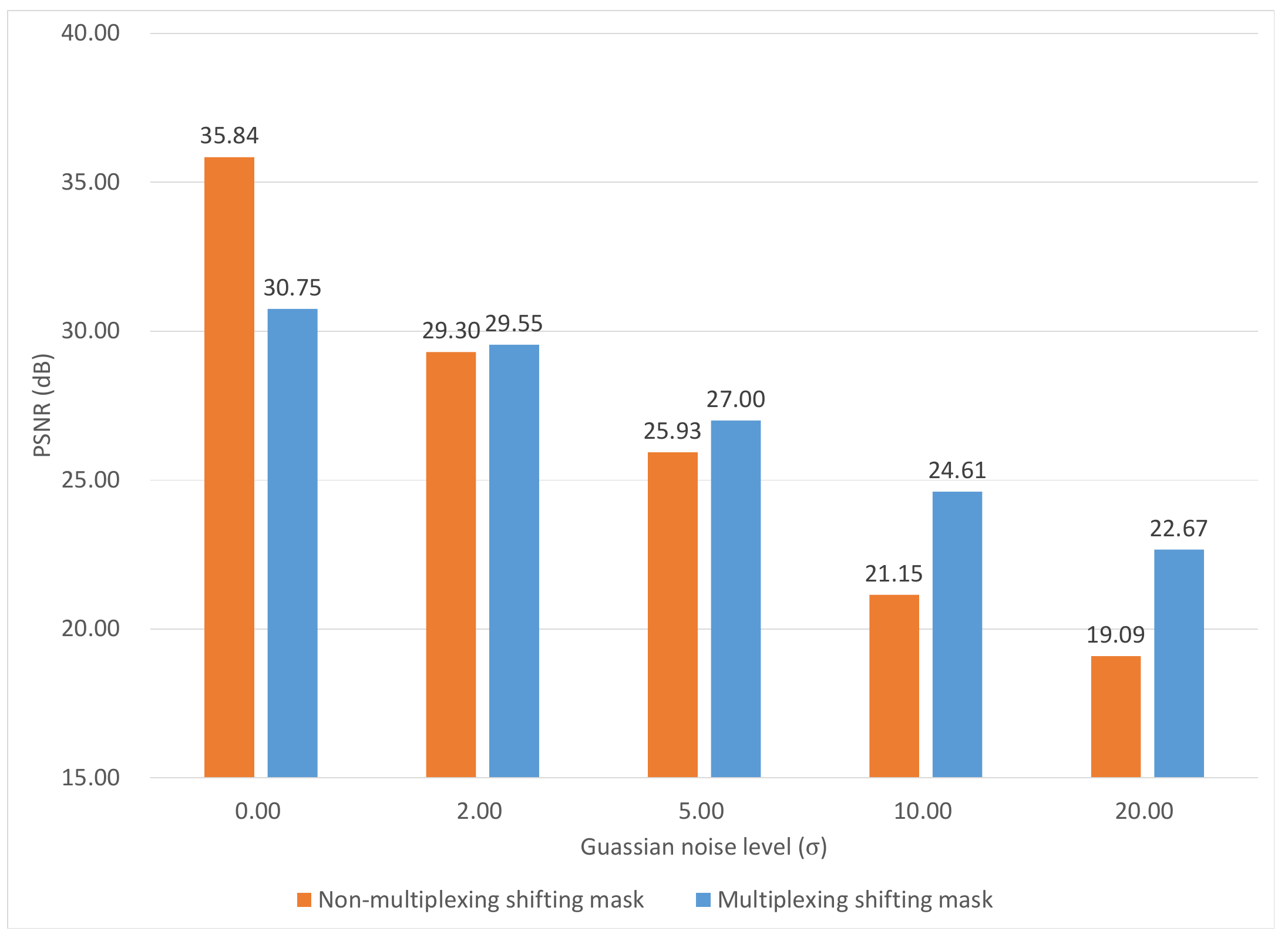}
\end{center}
\vspace{-3mm}
  \caption{Noise robustness comparison between multiplexed masks and non-multiplexed masks.}
\label{fig:exp2}
\end{figure}

\begin{figure*}
\begin{center}
    \includegraphics[width=.95\linewidth]{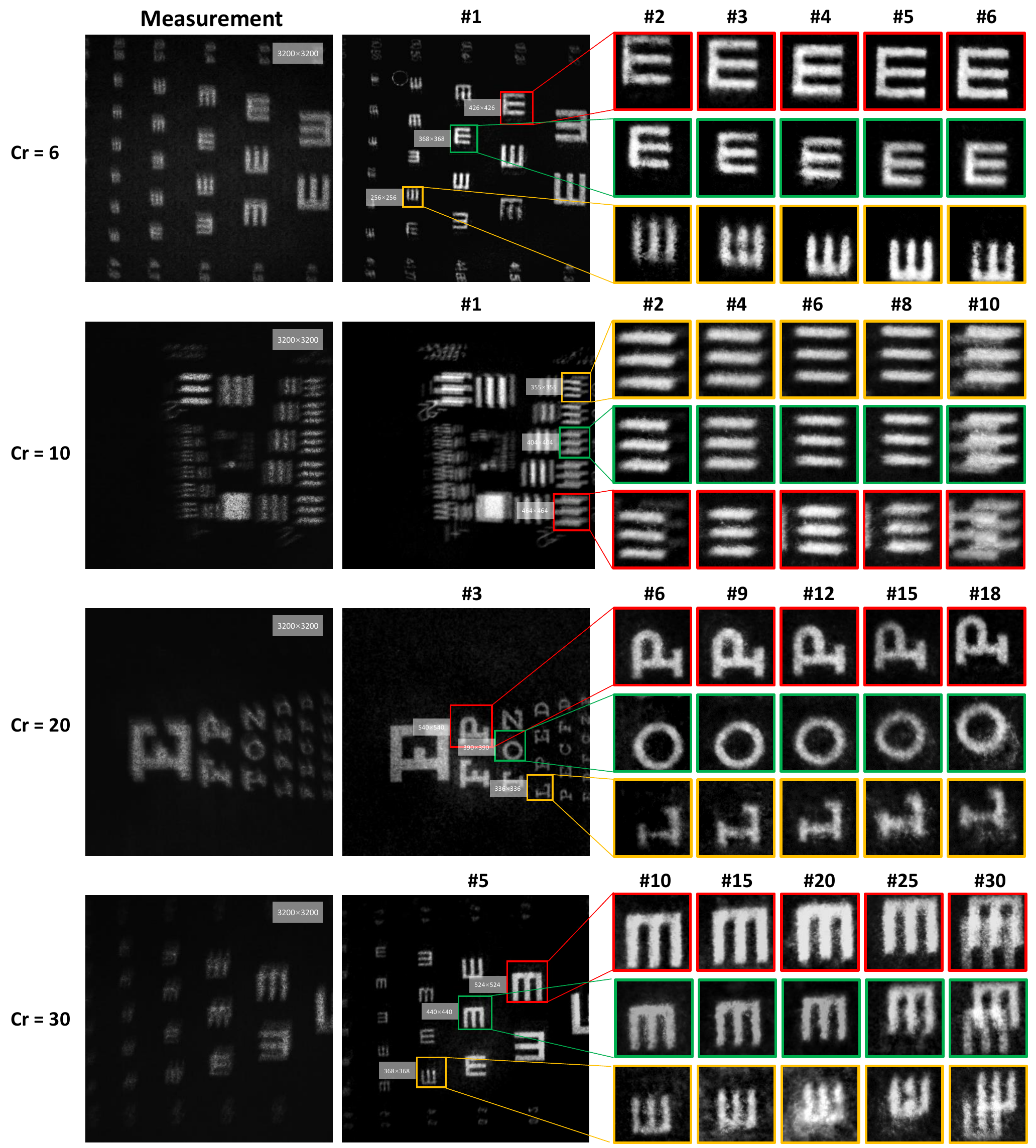}
\end{center}
\vspace{-3mm}
  \caption{Reconstruction results of PnP-TV-FastDVDNet on real data captured by our HCA-SCI system (Cr = 6, 10, 20 and 30). Note the full frames are of 3200\texttimes3200 and we plot small regions of sizes around 400\texttimes400 to demonstrate the high-speed motion.}
\label{fig:real_exp}
\vspace{-4mm}
\end{figure*}

\begin{figure*}
\begin{center}
    \includegraphics[width=.9\linewidth]{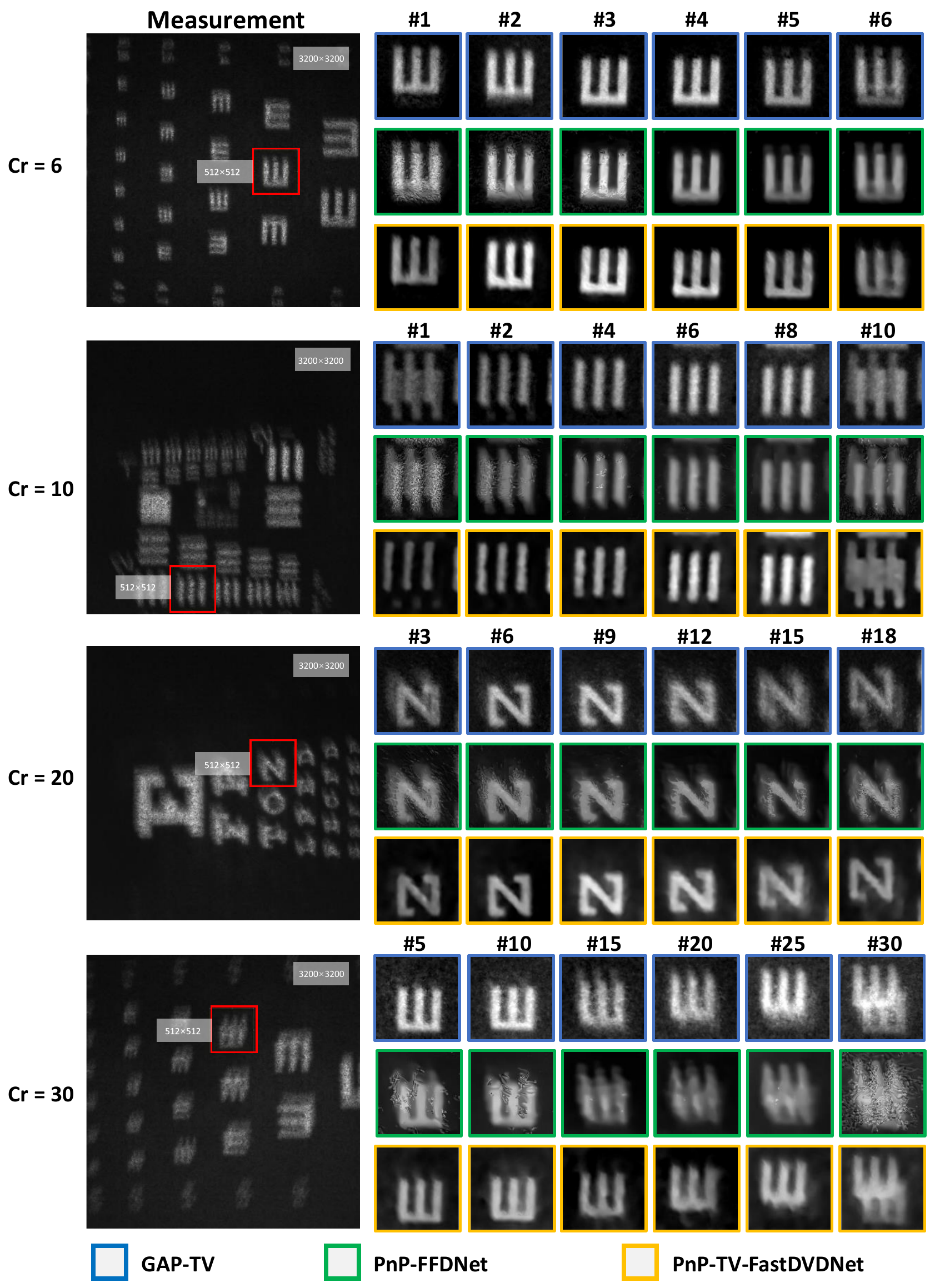}
\end{center}
\vspace{-3mm}
  \caption{Reconstruction comparison between GAP-TV, PnP-FFDNet and PnP-TV-FastDVDNet on real data captured by our HCA-SCI system (Cr = 6, 10, 20 and 30). Note the full frames are of 3200\texttimes3200 and we plot small regions of size of 512\texttimes 512 to demonstrate the high-speed motion. See Visualization 7 for the reconstructed videos.}
\label{fig:real_comp}
\vspace{-4mm}
\end{figure*}

\subsection{Validation of multiplexed shifting mask's noise robustness on simulation datasets}

As mentioned in Section \ref{sec: encoding_mask_gen}, our proposed HCA-SCI system leverages multiplexing strategy for the improvement of light throughput, which can thus gain a higher signal to noise ration (SNR) and enhance the system's robustness to noise in real-world applications. In this subsection, we conduct a series of experiments on simulation datasets containing different level of noise to compare the reconstruction performance of multiplexed shifting masks used in HCA-SCI and non-multiplexed ones. The $256 \times 256$ size dataset mentioned above are utilized here as the clean original frames. And we normalize the masks with a light throughput factor calculating from the proportion of the masks' active area with respect to the whole aperture. Zero-mean Gaussian noise with standard deviation ranging from 0 to 20 is added to the measurements (with values in $[0, 255]$) to simulate the real-world acquisition process.

The change of reconstruction PSNR over increasing noise levels is shown in Fig.~\ref{fig:exp2}. As can be seen from the figure, shifting masks with no multiplexing outperform the multiplexed ones when there is no noise, which is probably because the multiplexing operation brings in some correlation among the masks that is not desired for the encoding and reconstruction. However, as the noise increases, the superiority in SNR of the multiplexing scheme shows up, and the reconstruction PSNR of non-multiplexed shifting masks drops rapidly, while that of the multiplexed masks decreases more slowly and exceeds the PSNR of the non-multiplexed ones. In real scenes, noise is inevitable, and as the noise get larger, it will also impose significant impact on the reconstruction process until totally disrupt the reconstruction. So the multiplexing strategy leveraged in HCA-SCI can equip the physical system with more robustness in real applications, especially those at relatively large noise levels.

\subsection{Real data results}
We build a 10-mega-pixel snapshot compressive camera prototype illustrated in Fig.~\ref{fig:real_cover} for dynamic video recording. Empirically, the multiplexing patterns (shown in the second row of Fig.~\ref{fig:multiplex_pattern})  projected by the LCoS are designed to be rotationally symmetric which ensures the consistency in the final coding patterns and provides adequate incoherence among these masks. Before acquisition, we first calibrate the groups of coding masks with a Lambertian whiteboard placed at the objective plane, and each calibrated pattern is averaged over 50 repetitive snapshots to suppress the sensor noise. Then, during data capture, the camera operates at a fixed frame rate of 20 fps when Cr = 6, 10, and 20, providing reconstruction video frame rate of 120, 200 and 400 fps, respectively. For Cr = 30, the camera operates at 15 fps to extend the exposure time and provide a higher light throughput, which can reach a reconstruction frame rate up to 450 fps. We determine the spatial resolution of our system as 3200\texttimes3200 pixels. We thus have achieved the throughput of 4.6G voxels per second in the reconstructed video.

Three moving test charts printed on A4 papers are chosen as the dynamic objects. In Fig.~\ref{fig:real_exp}, we show the coded measurements and final reconstruction of the test charts. From which, one can see that the proposed HCA-SCI system and PnP-GAP-FastDVDNet reconstruction algorithm can effectively capture and restore the moving details of dynamic objects, which will be blurry when captured directly with conventional cameras. For the reconstruction of real data, we also find that, in some cases, the start and end frames tend to be blurry (refer to real data results of Cr = 10 and Cr = 30 in Fig.~\ref{fig:real_exp}), which might caused by the synchronization imperfection and initialization delay of the LCoS when switching between the projection sequences. 

We further compare the reconstruction of GAP-TV, PnP-FFDNet and PnP-TV-FastDVDNet on real datasets captured by our HCA-SCI system. From the reconstruction results shown in Fig.~\ref{fig:real_comp}, we can find that when Cr = 6, all these three algorithms can produce clear reconstruction frames with sharp details especially in the intermediate frames. But when Cr gets larger, the reconstructed frames of GAP-TV tend to be blurry and have more background noise. PnP-FFDNet will generate severe artifacts in the reconstructed frames and make the motion invisible. But our proposed PnP-TV-FastDVDNet can still reconstruct the motion details with a little increasing of noise in the background.


\section{Conclusion}
We have proposed a new computational imaging scheme capable of capturing 10 mega-pixel videos with low bandwidth and developed corresponding algorithms for computational reconstruction, providing a high throughput (up to 4.6G voxels per second) solution for high-speed, high-resolution videos. 

The hardware design bypasses the  pixel count limitation of available spatial light modulators via joint coding at aperture and close to image plane. The results demonstrate the feasibility of high throughput imaging under snapshot compressive sensing scheme and hold great potential for future applications in industrial visual inspection or multi-scale surveillance.

So far, the final reconstruction is limited to 450Hz since the hybrid coding scheme further decreases the light throughput to some extend compared with conventional coding strategies. In the future, a worthwhile extension is to introduce new photon-efficient aperture coding devices to raise the signal-to-noise ratio for coded measurements. Another limitation of the current system lies in the non uniformity of the encoding masks along the radial direction, which is a common problem for large-scale imaging systems due to non-negligible off-axis aberration. Considering the challenge for improving optical performance of existing physical components, designing novel algorithms which are capable of SCI reconstruction with non-uniform masks may be a economical and feasible way. Meanwhile, time-efficient reconstruction algorithms and feasible multiplexing pattern design methods for large-scale (like 10-mega pixel or even Giga pixel) SCI reconstruction is still an open challenge in the foreseeable future. Besides, extending proposed imaging scheme for higher throughput (e.g., Giga-pixel) or other dimensions (e.g., light field, hyperspectral imaging) may also be a promising direction.

\begin{backmatter}
\bmsection{Funding}
Ministry of Science and Technology of the People's Republic of China (Grant No. 2020AAA0108202); National Natural Science Foundation of China (Grant Nos. 61931012, 62088102).

\bmsection{Acknowledgments}
Zhihong Zhang and Chao Deng (marked with \dag ~in the author list) contributed equally to this paper.

\bmsection{Disclosures}
X. Y., Nokia (E).

\bmsection{Data Availability Statement}
Data underlying the results presented in this paper are not publicly available at this time but may be obtained from the authors upon reasonable request.

\end{backmatter}

\bibliography{sample}

\bibliographyfullrefs{sample}



\end{document}